\documentclass[11pt,conference,onecolumn]{IEEEtran}
\usepackage{cite}
\newtheorem{thm}{Theorem}
\newtheorem{cor}[thm]{Corollary}
\usepackage{amsmath,amssymb,amsfonts}

\long\def\symbolfootnote[#1]#2{\begingroup\def\thefootnote{\fnsymbol{footnote}}\footnote[#1]{#2}\endgroup}
\begin{document}
\title{Coding for Parallel Links to Maximize Expected Decodable-Message Value}

\author{
\authorblockN{Christopher~S.~Chang}
\authorblockA{Department of Electrical Engineering\\California Institute of Technology\\Pasadena, CA 91125, USA\\ Email: cswchang@caltech.edu}
\and
\authorblockN{Matthew~A.~Klimesh}
\authorblockA{Jet Propulsion Laboratory \\ California Institute of Technology\\Pasadena, CA 91109, USA\\ Email: Matthew.A.Klimesh@jpl.nasa.gov}}

\maketitle

\begin{abstract}
%``THIS PAPER IS ELIGIBLE FOR THE STUDENT PAPER AWARD''
Future communication scenarios for NASA spacecraft may involve multiple communication links and relay nodes, so that there is essentially a network in which there may be multiple paths from a sender to a destination. The availability of individual links may be uncertain. In this paper, scenarios are considered in which the goal is to maximize a payoff that assigns weight based on the worth of data and the probability of successful transmission. Ideally, the choice of what information to send over the various links will provide protection of high value data when many links are unavailable, yet result in communication of significant additional data when most links are available. Here the focus is on the simple network of multiple parallel links, where the links have known capacities and outage probabilities. Given a set of simple inter-link codes, linear programming can be used to find the optimal timesharing strategy among these codes. Some observations are made about the problem of determining all potentially useful codes, and techniques to assist in such determination are presented.
\end{abstract}

\section{Introduction}
%%%%%%%%%%%%%%%%%%%%%%%%%%%%%%%%%%%%%%%%%%%%%%%%%%%%%%%%%%%%%%%%%%%%%%%%%%%%%%%%%%%%%%%%%%%%%%%%%%%%%%%%%%%%%%%%%%
%
\symbolfootnote[0]{This research was carried out at the Jet Propulsion Laboratory, California Institute of Technology, under a contract with the National Aeronautics and Space Administration.}
When communicating through a noisy one-way communications link, it is well known that it is often practical to achieve a high level of reliability by protecting the communicated data with error-correcting codes, provided the attempted data rate is not too high. However, suppose that with some appreciable probability the communications link may fail for the entire duration of the communication attempt. Clearly, error-correcting codes are of no use in protecting against this type of link failure. But suppose there are more than one of these unreliable links. Then the communicated data can be protected somewhat with error-correcting codes applied between the multiple links. A trivial example of what we mean by this is if there are three links and identical data is sent on all links, thus protecting against failure of any two links (but not all three links).

More generally, we may want to send multiple messages with different worths (or priorities) from one point to another in a general network that contains unreliable links. Ideally, we would want to achieve the maximum available throughput at all times, in a way that: (1)~protects higher value data, and (2)~does not require prior knowledge of the network state. Usually this ideal goal will not be achievable, but we can consider tradeoffs. Roughly speaking, we would like to provide protection of higher value data when a large fraction of the links in a network are unavailable, and to achieve transmission of significant additional data when most links are working properly.

In this paper we restrict our attention to one simple type of network: a single source node and the single destination node connected by parallel unreliable links. We postulate that it is reasonable to model (or at least approximate) some communications scenarios as communications through this type of network. We develop and analyze a model whose main feature is communication through parallel unreliable links that do not change state (between working and nonworking) for the duration of the communication attempt. Some additional key features of our model are:
\begin{itemize}
\item a given link fails with some known probability, otherwise the link provides reliable communications with a known capacity; \item link failures are independent; \item the sender does not know the status of the links; \item there are some number of messages to be sent, and the messages have known worths (i.e., priorities or values) and sizes; \item the worth of a partial message is proportional to its size (i.e., partial credit is given for partial messages); and \item the payoff of a given link usage strategy is the expected total value of the messages successfully decoded.
\end{itemize}

The following (not entirely realistic) example scenario falls under our model. Consider a rover on Mars that, in a given period of time, can communicate to Earth in three ways: a direct-to-Earth link and two different relays through spacecraft orbiting Mars. It is assumed that these links have independent and non-negligible outage probabilities. 
It is also assumed that the resources (e.g., time and power) needed to utilize these links are small, so that the rover should always attempt to send information through all three routes. Then, given a list of messages and the values of their being received during this time period, we would like to find the best combinations of messages to send on each of the three links. Note that the separate links do not have to be used simultaneously to fit our model. In fact, it is possible to use our model to represent a single link used at different times, provided the outage probabilities can still be modeled as being independent.

We will present the model as though a given link capacity indicates the maximum amount of data that can be transmitted through the link (if it is up) during the communication attempt, and message sizes are also amounts of data on the same scale. Thus in a sense the communication attempt is an one-shot action. However, it would be equally valid to regard both capacities and message sizes as representing data rates per unit time. From this viewpoint the model would pertain to communication for an indefinite period of time, during which it is not possible to inform the sender which links are working.

Communication strategies for our model involve timesharing among relatively simple inter-link codes. Given a collection of candidate inter-link codes, the optimal timesharing proportions among them can be determined using linear programming. However, it appears to be difficult to determine the set of all ``useful'' inter-link codes for a given number of messages and links. We discuss this problem and describe an algorithm that can assist in determining all potentially useful inter-link codes that use all links equally, for a given number of messages and links.

In most communications scenarios previously studied, one is either interested in achieving arbitrarily small loss or error rates (in theoretical studies), or merely very small loss or error rates (in practical systems). The communication model described here is conceptually different in that it need not deal with small loss rates. 
Under our model we must accept that losses will occur and we measure performance in terms of the messages successfully communicated. This model may be appropriate for, say, a spacecraft that is capable of gathering much more information than it can transmit to Earth.

When generalized to general networks, our model can be thought of as a network coding problem. Most previously considered network coding problems fall into one of two categories: either they seek to maximize (a constant) throughput in a fixed network, or they seek robust communication at a constant rate in an unreliable network. Our model differs from both of these in that the throughput will depend on the network state even though a fixed coding scheme is used.

One somewhat related investigation of note is the Priority Encoding Transmission (PET) scheme of Albanese \textit{et al.}~\cite{albanese}. 
Also of note is the extension of Silva and Kschischang~\cite{silva} which shows how PET can be used in a network coding system. Like our model, the PET model includes a number of messages to be transmitted, and each message has an associated priority. Transmitted packets may be lost, but a given message can be recovered if a sufficient fraction of packets arrive successfully, where the fraction depends on the message's priority. One key difference between our model and the PET model is that under the PET model, there can be a large number of packets that may be received or lost independently, while under our model all packets sent on a given link are either lost or successfully received together, and there are a small number of different links. Our model includes outage probabilities, while under PET the only concern is the fraction of packets successfully received. Under PET all information is protected with (possibly trivial) maximum distance separable (MDS) codes. Under our model MDS codes may be used, but there is no reason to use block lengths larger than the number of links. Furthermore, we are concerned with partial decoding of codes, and this turns out to imply that non-MDS codes can be useful.

Indeed, even though we only need codes for correcting erasures and are primarily concerned with short block lengths, our particular model appears to make investigation of useful codes challenging. Not only are we interested in recovery of partial information when the whole codeword cannot be recovered, but the different codeword positions correspond to different links, and thus can have different erasure probabilities.

\subsection{Preliminaries} \label{preliminaries}
Let $N$ be the number of parallel links. For $i \in \{1,\ldots,N\}$, link $i$ has capacity $c_i$ and outage probability $p_i$ (equivalently, success probability $\bar{p}_i=1-p_i$). Let $M$ be the number of messages. For $j \in \{1,\ldots,M\}$, message $j$ has size $s_j$ and worth per unit size $\pi_j$. The units of the link capacities and message sizes are arbitrary (but are the same for all links and messages).

Informally, for each link $i$, the sender sends a stream of data that is a function of the $M$ messages and that is not longer than the link capacity $c_i$. The function can depend on all of the model parameters above. The receiver successfully receives the data on some subset of the links (and this subset is known to the receiver). The receiver then reconstructs the original messages to the extent possible from the data received. Let $R_j$ be a random variable indicating how many size units of message $j$ are recovered. The payoff from a communication attempt is the sum $\sum_{j=1}^M R_j \pi_j$. The payoff from a communication strategy is the expected value of this quantity.

It is implicit in this model that the message sizes are large and thus the messages can be split into pieces closely approximating any given fraction.

\subsection{Structure of the Paper} \label{structure}
In Section~\ref{section: General size and cap}, we propose a linear programming formulation to maximize the payoff for general capacities and general message sizes. In Section~\ref{section: Unit size and cap}, we consider unit capacity and unit message size case for which optimal code search methods and results are presented. Finally, Section~\ref{section: conclusions} concludes the paper.

\section{General Size and General Capacity Problem}\label{section: General size and cap}
In this section, we consider a general problem for different message sizes and link capacities. We assume that we can partition messages and code them together. Given a list of codes, we can determine how to optimally timeshare among them using linear programming.

It is convenient to regard communication strategies as consisting of (timesharing among) ``simple'' codes across the links. Consider the case of three links and two messages ($N=3$ and $M=2$). We label the messages $A$ and $B$. One possible code consists of sending a portion of message $A$ on all three links; we represent this code by $(A,A,A)$. Similarly, a code could consist of sending a portion of message $B$ on links 1 and 3; we represent this code by $(B,{-},B)$. Another possible code consists of sending a portion of message $A$ on link 1, an equal-sized portion of message $B$ on link 2, and the bitwise exclusive-or of these same portions on link 3; we represent this code by $(A,B,A+B)$, where the notation $A+B$ symbolizes addition in a finite field. A code can also use two different portions of the same message, as in $(A_1,A_2,A_1+A_2)$. We will avoid codes such as $(A,A,B)$ that can be achieved by timesharing simpler codes; in this case by equal parts of $(A,A,{-})$ and $({-},{-},B)$.

We assume that the message sizes and message worths are all given. We also assume that the link capacities and outage probabilities are given. Suppose our list of codes contains $n_c$ codes. Our objective is to find a column vector $\mathbf{z} = [z_1,\ldots,z_{n_c}]^T$ that describes how much we use each code.

The codes are described with an $N \times n_c$ matrix $\mathbf{K}$ that tells how much the codes use the links, an $M \times n_c$ matrix $\mathbf{L}$ that tells the message content of the codes, and an $n_c$ element column vector $\mathbf{v}$ that gives the expected payoffs from the codes.

More specifically, in $\mathbf{K}=[k_{i,k}]$ the entry $k_{i,k}$ is the usage of link $i$ by one unit of code $k$. In $\mathbf{L}=[\ell_{j,k}]$ the entry $\ell_{j,k}$ is the amount of message $j$ sent with one unit of code $k$. And in $\mathbf{v} = [v_1,\ldots,v_{n_c}]^T$ the entry $v_k$ is the payoff (in expected received value) from one unit of code $k$; note that $v_k$ is a function of the message worths and the link outage probabilities as well as of the properties of code $k$.

Observe that each code is described by an entry in $\mathbf{v}$ and a column in each of $\mathbf{K}$ and $\mathbf{L}$. As an example, consider the codes $(A,B,A+B)$ and $(A_1,A_2,A_1+A_2)$ for the case of two messages and three links. We can describe $(A,B,A+B)$ by the columns $(1,1,1)$ and $(1,1)$ in $\mathbf{K}$ and $\mathbf{L}$ respectively. Note that the unit size is arbitrary here; we still describe the same code if we scale both columns by an arbitrary factor (and the entry in $\mathbf{v}$ would need to be scaled by the same factor). The code $(A_1,A_2,A_1+A_2)$ can be described by $(1,1,1)$ and $(2,0)$.

The linear programming problem is used to find a column vector $\mathbf{z}$ which maximizes $\mathbf{v^T z}$ subject to $\mathbf{z} \geq 0$ and $\mathbf{Kz} \leq \mathbf{c}$ and $\mathbf{Lz} \leq \mathbf{s}$, where $\mathbf{c} = (c_1,\ldots,c_N)$ is the link capacity list and $\mathbf{s} = (s_1,\ldots,s_M)$ is the message size list. Now, the linear programming formulation is as follows:
\begin{equation}\label{linprog for coding}
\begin{split}
\mbox{find } \mathbf{z} \mbox{ to maximize} & \quad \mathbf{v}^T \mathbf{z}\\
\mbox{subject to}& \quad \mathbf{K} \mathbf{z} \le [c_1,\ldots,c_N]^T \\ % number is here
& \quad \mathbf{L} \mathbf{z} \le [s_1,\ldots,s_M]^T \\ % number is here
& \quad \mathbf{z} \geq 0
\end{split}
\end{equation}

As a concrete example, consider the case of $2$ messages and $3$ parallel links. This case is simple enough that we can write down a list of $17$ codes that (probably) includes all the codes that we need to consider:
\[ \begin{split} 
%\\
 \{ (A,&{-},{-}),\, ({-},A,{-}),\, ({-},{-},A),\, (A,A,{-}),\, (A,{-},A),\\
  &(A,A,{-}),\, (A,A,A),\, (A_1,A_2,A_1+A_2),\, (B,{-},{-}), \\
  &({-},B,{-}),\, ({-},{-},B),\, (B,B,{-}),\, (B,{-},B),\, (B,B,{-}), \\
  &(B,B,B),\, (B_1,B_2,B_1+B_2),\, (A,B,A+B) \} \end{split} \]

Note that we have assumed the links are indexed so that $p_1 \leq \cdots \leq p_N$ (decreasing reliability) and that the messages are indexed in order of decreasing value (per unit size), $\pi_1 \geq \cdots \geq \pi_M$. With these assumptions it is easily verified that we do not need to consider any other permutations of symbols in the codes $(A_1,A_2,A_1+A_2)$, $(B_1,B_2,B_1+B_2)$, and $(A,B,A+B)$; for example the code $(A,B,A+B)$ is always at least as good as $(B,A+B,A)$.

For this list of codes we have the following:
\begin{equation*}
\begin{split}
\mathbf{K} =& %\scriptsize
  \left(
\begin{array}{c}
1	\ \ 0 \ \ 0	\ \ 1	\ \ 1	\ \ 0	\ \ 1	\ \ \tfrac{1}{2}	\ \ 1 \ \ 0	\ \ 0	\ \ 1	\ \ 1	\ \ 0	\ \ 1	\ \ \tfrac{1}{2}	\ \ 1\\
0	\ \ 1	\ \ 0	\ \ 1	\ \ 0	\ \ 1	\ \ 1	\ \ \tfrac{1}{2}	\ \ 0 \ \ 1	\ \ 0	\ \ 1	\ \ 0	\ \ 1	\ \ 1	\ \ \tfrac{1}{2}	\ \ 1\\
0	\ \ 0	\ \ 1	\ \ 0	\ \ 1	\ \ 1	\ \ 1	\ \ \tfrac{1}{2}	\ \ 0 \ \ 0	\ \ 1	\ \ 0	\ \ 1	\ \ 1	\ \ 1	\ \ \tfrac{1}{2}	\ \ 1
\end{array} \right) \\
\mathbf{L} =& %\scriptsize
  \left( \begin{array}{c}
1 \ \ 1 \ \ 1 \ \ 1 \ \ 1 \ \ 1 \ \ 1 \ \ 1 \ \ 0 \ \ 0 \ \ 0 \ \ 0 \ \ 0 \ \ 0 \ \ 0 \ \ 0 \ \ 1\\
0 \ \ 0 \ \ 0 \ \ 0 \ \ 0 \ \ 0 \ \ 0 \ \ 0 \ \ 1 \ \ 1 \ \ 1 \ \ 1 \ \ 1 \ \ 1 \ \ 1 \ \ 1 \ \ 1 
\end{array} \right)
\end{split}\end{equation*}
\[%\scriptsize
\mathbf{v}:\left\{ \begin{array}{rcl}
\left[v_1, v_9\right] & = & \left[\pi_A, \pi_B\right] \ \bar{p}_1\\
\left[v_2, v_{10}\right] & = & \left[\pi_A, \pi_B\right] \ \bar{p}_2\\
\left[v_3, v_{11}\right] & = & \left[\pi_A, \pi_B\right] \ \bar{p}_3\\
\left[v_4, v_{12}\right] & = & \left[\pi_A, \pi_B\right](\bar{p}_1 + \bar{p}_2 - \bar{p}_1 \bar{p}_2)\\
\left[v_5, v_{13}\right] & = & \left[\pi_A, \pi_B\right](\bar{p}_1 + \bar{p}_3 - \bar{p}_1 \bar{p}_3)\\
\left[v_6, v_{14}\right] & = & \left[\pi_A, \pi_B\right](\bar{p}_2 + \bar{p}_3 - \bar{p}_2 \bar{p}_3)\\
\left[v_7, v_{15}\right] & = & \left[\pi_A, \pi_B\right](\bar{p}_1 + \bar{p}_2 + \bar{p}_3 - \bar{p}_1 \bar{p}_2 -
\bar{p}_1 \bar{p}_3 - \bar{p}_2 \bar{p}_3 + \bar{p}_1 \bar{p}_2 \bar{p}_3)\\
\left[v_8, v_{16}\right] & = & \left[\pi_A, \pi_B\right]( 0.5 \bar{p}_1 (1-\bar{p}_2) (1-\bar{p}_3)  + 0.5 \bar{p}_2
(1-\bar{p}_1) (1-\bar{p}_3) \\ && \qquad \qquad + \bar{p}_1 \bar{p}_2 + \bar{p}_1 \bar{p}_3 + \bar{p}_2 \bar{p}_3 - 2 \bar{p}_1 \bar{p}_2 \bar{p}_3)\\
v_{17} & = & \pi_A   \bar{p}_1   (1-\bar{p}_2) (1-\bar{p}_3) + \pi_B 
\bar{p}_2  (1-\bar{p}_1) (1-\bar{p}_3) \\ && + \left(\pi_A+\pi_B \right)(\bar{p}_1 \bar{p}_2 + \bar{p}_1 \bar{p}_3 + \bar{p}_2 \bar{p}_3 -
2 \bar{p}_1 \bar{p}_2 \bar{p}_3)   \end{array} \right.
\]

For example, $v_4 = \pi_A(\bar{p}_1 + \bar{p}_2 - \bar{p}_1 \bar{p}_2)$ because when one unit of the fourth code is used, one unit of message $A$ can be decoded (with worth $\pi_A$) whenever at least one of the first two links are up (which happens with probability $\bar{p}_1 +
\bar{p}_2 - \bar{p}_1 \bar{p}_2$).

We solved the linear programming problem for a number of randomly generated scenarios (random message sizes and worths, and random link capacities and probabilities). We found that for each of the $17$ codes, there were scenarios in which the code was needed in the optimal solution.

The linear programming method clearly has limitations for our problem. 
For one thing, making the list of candidate codes looks to be a very complicated problem in general. In addition, the number of candidate codes grows at least exponentially with the number of links, since there are already $2^N-1$ possibilities just for repetition codes for the first message. However, the linear programming method appears to be viable for small numbers of links, which is a case that may be of interest for some applications.

\section{Unit Size and Unit Capacity Problem}\label{section: Unit size and cap}
In this section, we consider a special case of the problem with unit sizes and unit capacities. The motivation for this is try to characterize, or at least be able to generate, complete sets of candidate codes for the general size, general capacity case. And note that the coding used in this section corresponds to the ``scalar linear combination'' in the literature of network coding.

To determine the best coding scheme (i.e., how to choose the best code for a given parameters), we consider all possible candidates, and through a number of random trials, we can find the best code and related rules. If allowing coding across the links, however, we have enormous number of possible codes even if we consider only linear combination for coding. Therefore, to pick the best coding scheme, generating all possible codes and then comparing their performances are almost intractable even for a small number of links case. To deal with this problem, we take advantage of a connection between entropy and matroids.

For our purposes, we need only the following characterization of matroids, which is Corollary~1.3.4 in~\cite{oxley}:
\begin{cor}\label{col_1_3_4}
Let $E$ be a set. A function $r : 2^E \rightarrow \mathbb{Z}^+ \cup \left\{ 0\right\} $ is the rank function of a matroid on $E$ if and only if $r$ satisfies the following conditions:
\begin{itemize}
\item[(R1)] If $X \subseteq E$. then $0 \le r(X) \le \left|X \right|$.
\item[(R2)] If $X \subseteq  Y \subseteq E$, then $r(X) \le r(Y)$.
\item[(R3)] If $X$ and $Y$ are subsets of $E$, then $r(X \cup Y) + r(X \cap Y) \le r(X) + r(Y)$.
\end{itemize}
\end{cor}

Now, we consider the connection between matroids and codes in the following subsection.

\subsection{Codes and Matroids}
Suppose we have a code $C$ for $M$ messages and $N$ links, say $C=(X_1,\ldots,X_N)$ where each $X_i$ is generated according to $X_i=f_i(\mathbf{M})$ with $\mathbf{M}=(M_1,\ldots,M_M)$ and the $M_j$ are messages of equal size. We assume sizes are normalized so that the message sizes are all 1.

Suppose now that the messages, $M_1,\ldots,M_M$, are each random variables, independent and uniformly distributed on their possible values. Then $X_1,\ldots,X_N$ are also random variables since they are functions of $\mathbf{M}$. Let $U_1 = \{M_1,\ldots,M_M\}$, $U_2 = \{X_1,\ldots,X_N\}$, and $U = U_1 \cup U_2$. We consider the joint entropies of subsets of $U$. For convenience, we normalize the entropies so that any message $M_j$ has unit entropy, i.e., $H(M_j)=1$.

Now define $f:2^U \rightarrow [0,\infty)$ by $f(S)=H(S)$, where $2^U$ is the power set of $U$ and $H(S)$ is the joint entropy of the members of $S$ (with $H(\varnothing)=0$). As is well known, the nonnegativity of conditional mutual information implies that this function is submodular, meaning $f(S_1 \cup S_2)+f(S_1 \cap S_2) \leq f(S_1)+f(S_2)$. Clearly $f$ is also monotone, meaning $f(S_1) \leq 
f(S_2)$ whenever $S_1 \subseteq S_2$. These properties, along with $f(\varnothing)=0$, mean $f$ is by definition a \emph{polymatroid function} \cite{Fuj05,Mat07,Dou07}. The fact that entropy is a polymatroid function is well known; see e.g.,~\cite{Mat07,Dou07}.

For simplicity, it seems reasonable to consider only codes for which $f$ defined in this way is integer-valued. In any case, if a code produces an $f$ that takes on noninteger but rational values, we can choose a new normalization of the message entropies to make the corresponding $f$ take on only integer values, then the messages can be subdivided so that they have unit entropy again. The resulting code would be a code on a larger number of messages but would be essentially equivalent to the original code.

To further simplify, we will restrict ourselves to codes for which all $X_i$ also have unit entropy (implying that the encoded symbols are ``messages'' of unit length). However, we think it is likely that there are codes that are ``useful'' and do not satisfy this condition.

With this further simplification the function $f$ must be the rank function of a matroid (see \cite{oxley,Mat07}), as it is integer-valued and satisfies $f(S) \leq |S|$.

Thus any code that satisfies our conditions has a corresponding matroid. However, the converse probably does not hold: it is likely that there are matroids that cannot be produced from a code as above. This is because it is known that there are matroids that are non-entropic, meaning there does not exist an ensemble of random variables with joint entropies corresponding to the matroids's rank function. In fact it is known that there are matroids that are not asymptotically entropic, which, loosely speaking, means they cannot be approximated closely by entropic polymatroids. See \cite{Mat07} for a more precise definition of asymptotically entropic matroids and an example of a matroid that is not asymptotically entropic (the V\'{a}mos matroid).

Our interest in matroid rank functions for codes stems from two observations. First, the matroid rank function contains complete information about the performance of the code. Second, it appears to be 
much easier to systematically generate all matroid rank functions for a given number of messages and links than it is to generate all codes (or code properties).

We consider obtaining information about the performance of the code from the matroid rank function. First note that we require $f(S)=|S|$ when $S \subseteq U_1$, since the messages are independent. We also require $f(\{X_i\})=1$. The properties of matroid rank functions imply that if $S \subseteq U_2$ and $M_j \in U_1$, then either $f(S \cup \{M_j\}) = f(S)$ or $f(S \cup \{M_j\}) = f(S)+1$. The latter implies $M_j$ is independent of $S$, and so cannot be determined from $S$. The former implies $M_j$ is completely determined from $S$ and so $M_j$ can be recovered when the links corresponding to $S$ are up.

We remark that our hypothesis that the messages are random and independent is only a tool for analysis; we do not require the messages to be random in an actual communications system. Clearly, if we want to be able to decode some $M_j$ from a subset $S$ of the code symbols, it must be necessary for this to be possible when the messages are random and independent. And if it is possible in that case, then clearly $M_j$ must be deterministically obtainable from $S$ no matter what the messages are.

As we mentioned earlier, some matroid rank functions may not correspond to any realizable code. However, if a code produces performance better or equal to that corresponding to any matroid rank function, we can still conclude the code must be optimal.

%%%%%%%%%%%%%%%%%%%%%%%%%%%%%%%%%%%%%%%%%%%%%%%%%%%%%%%%%%%%%%%%%%%%%%%%%%%%%%%%%%%%%%%%%%%%%%%%%%%%%%%%%%
\subsection{Automated Process: Generating Matroids and Calculating Performance Metrics}\label{section: automated_process}
When we have $M$ messages and $N$ links, we consider a matroid with $(M+N)$ elements. By enumerating candidates for rank functions, we can count all possible matroids. Recall that the maximum rank is $M$ and any singleton element has rank $1$. Therefore, without considering conditions for rank functions in Corollary~\ref{col_1_3_4}, there are $M^{2^{(M+N)}-1}$ possible cases.

To efficiently generate all rank functions with given parameters, we use a backtracking algorithm~\cite{brassard}. Whenever we assign a possible value ($1,2,\ldots,M$) to an unassigned variable (rank function), we check the validity of that assignment by checking the conditions in Corollary~\ref{col_1_3_4}.

Once we have a full list of all possible matroids, we can calculate the payoff for each matroid (each case) with an automated process. Recall that if $r(S) = r(S \cup M_j)$, then $M_j$ is decodable from a set of links, $S$. When we have $N$ links, we need to compare $2^N-1$ combinations of links (excluding the empty set). For each combination, we check if the message $M_j$ is decodable with that combination or not, using $r(S) = r(S \cup M_j)$. For each combination, we have the corresponding probability that is weighted by the worth of the message and then added to the payoff.

\subsection{Systematic Codes}
If the performance of a given code is never better than the performance of another code, then the former code can be eliminated from consideration. As an example, consider the codes $(A,B,A+B,A+C)$ and $(A,B,A+B,C)$. In $(A,B,A+B,A+C)$, message $C$ appears only once (link~$4$), and that is in combination with message $A$. Thus it is necessary but not sufficient for link~$4$ to be up to recover message $C$, and it can be verified that $A+C$ is never useful for recovering any other message. Therefore we cannot do any worse by replacing $A+C$ with $C$. More generally, by this reasoning we can eliminate from consideration any code for which a message is only involved in one code position, and the message is combined with one or more other messages there.

One might conjecture that this idea could be generalized further and we need only consider \emph{systematic} codes, which for our purposes are codes in which any message involved in the code is sent directly (not combined with other messages) on some link. (For example, $( A,A,A)$ and $( A,B,A+B)$ are systematic codes under this definition.)  However, perhaps surprisingly, this conjecture appears to be false: for $5$ links and $3$ messages, our preliminary results indicate that some permutations of the codes $(A,B,A+B,A+C,B+\alpha C)$, $(A,C,A+B,A+C,B+\alpha C)$ can be optimal. (Here the codes are in a non-binary field and $\alpha$ is not $1$.) These potential counterexamples were arrived at by enumerating what we think is all codes that could conceivably be optimal, then generating random scenarios (worths and outage probabilities) and checking which code gives the best payoff. This will be considered in more depth in Section~\ref{section: simulation}.

When comparing the payoffs for different cases using the automated process described in Section~\ref{section: automated_process}, we are more interested in if the best case is a systematic code, than the actual payoff value. Therefore, we need another automated process for checking if the code is systematic or non-systematic.
So, we have the following criterion:\\

\begin{itemize}
\item[(C1)] For all messages such that $r(U_2) = r(U_2 \cup M_j)$ (meaning that each message $M_j$ is included in the code), there exists at least one link $i$ such that $r(i) = r(i \cup M_j)=1$.\\
\end{itemize}

If all messages in the code satisfy the condition (C1), then the code is declared as a systematic code.

%%%%%%%%%%%%%%%%%%%%%%%%%%%%%%%%%%%%%%%%%%%%%%%%%%%%%%%%%%%%%%%%%%%%%%%%%%%%%%%%%%%%%%%%%%%%%%%%%%%%%%%%%%%%%%%%%%%%%%%
\subsection{Random Simulation Results}\label{section: simulation}
We have searched the optimum codes for $3$ messages with $4$ and $5$ links. For $4$ links case, we construct a full list of matroids with which we simulated random trials. When we consider all possible coding schemes with random outage probability and random worth in the range of $[1, 100]$, it would be a counterexample to the conjecture if we obtain a non-systematic code that has the best payoff among all coding schemes. However, while simulating a large number of trials, we have never encountered a non-systematic as the best code for any instance. Therefore, this simulation results support the conjecture in the case of $3$ messages with $4$ links.

For $5$ links case, we do not have a full list of matroids yet. Therefore, we consider a partial list with all possible code candidates from which we removed the obviously worse code combinations (i.e., we enumerate some coding schemes which seem like they have some chance to be optimal). In contrast to $4$ links case, we found a surprising result in this case: non-systematic codes such as $(A,B,A+B,A+C,B+\alpha C)$ and $(A,C,A+B,A+C,B+\alpha C)$ ($\alpha \neq 1$) can be optimal in some instances. With those counterexamples, we surmise that the conjecture is not valid anymore. Furthermore, if the conjecture is not valid for unit capacity and unit size case, it cannot be generalized to different capacity and different size case, either. 

%%%%%%%%%%%%%%%%%%%%%%%%%%%%%%%%%%%%%%%%%%%%%%%%%%%%%%%%%%%%%%%%%%%%%%%%%%%%%%%%%%%%%%%%%%%%%%%%%%%%%%%%%%%%%%%%%%%%%%%

\section{Conclusions}\label{section: conclusions}

%%%%%%%%%%%%%%%%%%%%%%%%%%%%%%%%%%%%%%%%%%%%%%%%%%%%%%%%%%%%%%%%%%%%%%%%%%%%%%%%%%%%%%%%%%%%%%%%%%%%%%%%%%%%%%%%%%%%%%%

We have presented a problem formulation for transmitting messages with different worths through an unreliable network. As a fundamental network model, we have considered a unicast communication connected by multiple parallel links. 

We suggested a linear programming formulation that allows us to determine what combinations of messages to use across the multiple channels among a precomputed library of simple combinations of messages. We also described an algorithm for assisting in finding viable combinations of messages to populate this precomputed library.

\section{Acknowledgments}
The author would like to thank Tracey Ho for several insightful discussions.

\bibliographystyle{IEEEtran}
\bibliography{IEEEabrv,networkcoding}

\end{document}